\documentclass[preprint]{aastex61}
\received{May 15, 2017}
\revised{May 30, 2017}
\accepted{June 2, 2017}
\submitjournal{ApJL}

\begin{document}

\def\aj{Astronomical Journal,~}
\def\actaa{Acta Astronomica,~}
\def\araa{Annual Review of Astronomy and Astrophys,~}
\def\apj{Astrophysical Journal,~}
\def\apjl{Astrophysical Journal, Letters,~}
\def\apjs{Astrophysical Journal, Supplement,~}
\def\ao{Applied Optics,~}
\def\apss{Astrophysics and Space Science,~}
\def\aap{Astronomy and Astrophysics,~}
\def\aapr{Astronomy and Astrophysics Reviews,~}
\def\aaps{Astronomy and Astrophysics, Supplement,~}
\def\azh{Astronomicheskii Zhurnal,~}
\def\baas{Bulletin of the AAS,~}
\def\caa{Chinese Astronomy and Astrophysics,~}
\def\cjaa{Chinese Journal of Astronomy and Astrophysics,~}
\def\icarus{Icarus,~}
\def\jcap{Journal of Cosmology and Astroparticle Physics,~}
\def\jrasc{Journal of the RAS of Canada,~}
\def\memras{Memoirs of the RAS,~}
\def\mnras{Monthly Notices of the RAS,~}
\def\na{New Astronomy,~}
\def\nar{New Astronomy Review,~}
\def\pra{Physical Review A: General Physics,~}
\def\prb{Physical Review B: Solid State,~}
\def\prc{Physical Review C,~}
\def\prd{Physical Review D,~}
\def\pre{Physical Review E,~}
\def\prl{Physical Review Letters,~}
\def\pasa{Publications of the Astron. Soc. of Australia,~}
\def\pasp{Publications of the ASP,~}
\def\pasj{Publications of the ASJ,~}
\def\rmxaa{Revista Mexicana de Astronomia y Astrofisica,~}
\def\qjras{Quarterly Journal of the RAS,~}
\def\skytel{Sky and Telescope,~}
\def\solphys{Solar Physics,~}
\def\sovast{Soviet Astronomy,~}
\def\ssr{Space Science Reviews,~}
\def\zap{Zeitschrift fuer Astrophysik,~}
\def\nat{Nature,~}
\def\iaucirc{IAU Cirulars,~}
\def\aplett{Astrophysics Letters,~}
\def\apspr{Astrophysics Space Physics Research,~}
\def\bain{Bulletin Astronomical Institute of the Netherlands,~}
\def\fcp{Fundamental Cosmic Physics,~}
\def\gca{Geochimica Cosmochimica Acta,~}
\def\grl{Geophysics Research Letters,~}
\def\jcp{Journal of Chemical Physics,~}
\def\jgr{Journal of Geophysics Research,~}
\def\memsai{Mem. Societa Astronomica Italiana,~}
\def\nphysa{Nuclear Physics A,~}
\def\physrep{Physics Reports,~}
\def\physscr{Physica Scripta,~}
\def\planss{Planetary Space Science,~}
\def\procspie{Proceedings of the SPIE,~}

%
%


\title{Reconciling the dawn-dusk asymmetry in Mercury's exosphere with the micrometeoroid impact directionality }

%
%




\correspondingauthor{Petr Pokorn\'{y}}
\email{petr.pokorny@nasa.gov}

\author{Petr Pokorn\'{y}}
\affiliation{Department of Physics, The Catholic University of America, Washington, DC 20064, USA}
\affiliation{Space Weather Laboratory, Code 674, NASA Goddard Space Flight Center, Greenbelt, MD 20771}

\author{Menelaos Sarantos}
\affiliation{Geospace Physics Laboratory, Code 673, NASA Goddard Space Flight Center, Greenbelt, MD 20771}

\author{Diego Janches}
\affiliation{Space Weather Laboratory, Code 674, NASA Goddard Space Flight Center, Greenbelt, MD 20771}

%
%
%



%

%
%


\begin{abstract}
Combining dynamical models of dust from Jupiter Family Comets and Halley-type Comets, we demonstrate that the seasonal variation of the dust/meteoroid environment at Mercury is responsible for producing the dawn-dusk asymmetry in Mercury's exosphere observed by the MESSENGER spacecraft. Our latest models, calibrated recently from ground-based and space-borne measurements, provide unprecedented statistics that enable us to study the longitudinal and latitudinal distribution of meteoroids impacting Mercury's surface. We predict that the micrometeoroid impact vaporization source is expected to undergo significant motion on Mercury's surface towards the nightside during Mercury's approach to aphelion and towards the dayside when the planet is approaching the Sun.

\end{abstract}


\section{Introduction}

Micrometeoroid impact vaporization has long been considered to be an important source of metals in Mercury's exosphere. This link has been strengthened recently when measurements by the UltraViolet and Visible Spectrometer (UVVS) instrument on-board of the NASA MErcury Surface, Space ENvironment, GEochemistry and Ranging (MESSENGER) spacecraft provided measurements of unprecedented quality of Mercury's exosphere, which permit the study of the seasonal variations of exotic metals like Mg \citep{Merkel_etal_2017} and Ca \citep{Burger_etal_2014}.  Measurements of such exospheric metals yield - by way of scale height measurements - source process temperatures that appear consistent with meteoroid impacts vaporizing from Mercury's soil both atoms and molecules which quickly dissociate. Furthermore, the dependence of the inferred source rate with Mercury's True Anomaly Angle (TAA) approximately matches predictions of how the impact vapor varies with heliocentric distance \citep{Killen_Hahn_2015}. Thirdly, the morphology of Mg and Ca emissions is inconsistent with ion sputtering, the other process which can release energetic atoms from the soil: both Ca and Mg exhibit a dawn and/or morning enhancement accompanied by a pronounced dawn-dusk asymmetry~\citep{Mcclintock_etal_2009, Burger_etal_2014, Merkel_etal_2017}, which has by analogy to Earth been attributed to the directionality of micrometeoroid impact. However, this last point has not yet been reproduced by models of Mercury's impactors, and this is the main motivation for our paper.

Properties of the meteoroid environment around Mercury have been investigated by several groups in the last decade. \citet{Marchi_etal_2005} presented an estimate of the meteoroid flux on Mercury for meteoroids with sizes $>1~\mathrm{cm}$ originating from asteroids and Jupiter-Family Comets (JFCs). \citet{Borin_etal_2009} focused on dynamics of smaller meteoroids with sizes between $5\mu\mathrm{m}$ and $1000\mu\mathrm{m}$ from asteroids and provided  impactor velocity distributions similar to \citet{Cintala_1992} but with much higher impactor fluxes. \citet{Borin_etal_2016a} reduced these estimated fluxes by adopting a revised dust particle flux impacting the Earth, and added meteoroids from JFCs in \citet{Borin_etal_2016b}, proposing that asteroids and JFCs likely have equal importance at Mercury. More recently, \citet{Killen_Hahn_2015} used the velocity distributions reported by \citet{Cintala_1992} and combined them with analytically prescribed meteoroid disks to fit the variations of Ca vapor rates with changing TAA. These investigations focused mainly on quantifying the expected impact vaporization rates, finding them to be sufficient to produce the observed exosphere, whereas in this paper we focus on the directionality of impacts.

Our knowledge about the impact flux asymmetry on Mercury's surface has been limited thus far. The only evaluation of diurnal asymmetries of the meteoroid flux on Mercury's surface was given by \citet{Marchi_etal_2005}, who noted that large impactors ($>13~\mathrm{cm}$) preferentially impact the dayside. Similarly limited is our knowledge of the latitudinal dependence of impactors:  \citet{Borin_etal_2016a} studied the latitudinal distribution of impacts of meteoroids smaller than 1 cm, finding the concentration near the equator, however, the low impact statistics forced authors to sum impacts from all TAAs.
In this paper we eliminate the statistical limitations of previous works and add previously untraced dust particles from Halley Type Comets (HTCs) to present the first evaluation of micrometeoroid impact asymmetries in the size interval $10-200~\mu\mathrm{m}$, relevant to the daily replenishment of the exosphere, as a function of local time, latitude and Mercury TAA.

\section{Meteoroid Models}
We modeled the impact on Mercury's surface of two meteoroid populations produced by JFCs and HTCs. The dynamics and evolution of meteoroids from JFCs were extensively modeled and studied by \citet{Nesvorny_etal_2010} and then revised using more observational constraints by \citet{Nesvorny_etal_2011JFC}. 
In this work, we use models from \citet{Nesvorny_etal_2011JFC} who studied 29 different sizes of meteoroids. 
For dust originating from HTCs, we use results reported by \citet{Pokorny_etal_2014}, who studied the evolution of 12 different sizes of meteoroids. In this study we omit other meteoroid producing populations for the following reasons: 1) asteroidal meteoroids do not significantly contribute to the total budget of meteoroids in the inner Solar System \citep{Nesvorny_etal_2011JFC}. Furthermore, due to their nearly circular orbits, they are subject to significant collisional fading before reaching Mercury's orbit, an effect which was not included in the model of \citet{Borin_etal_2009,Borin_etal_2016a}; 2) Oort Cloud Comet (OCC) meteoroids have been so far thoroughly modeled only by \citet{Nesvorny_etal_2011OCC} whose model currently suffers from low statistics at Earth and more simulations are required to properly model this component; 3) Kuiper Belt Objects (KBOs) meteoroids might reach the inner Solar System, however their significance is negligible in this region \citep{Nesvorny_etal_2010, Poppe_2016}. 

The meteoroid populations considered here are expected to be the vast majority of the sporadic meteoroid environment in the inner Solar System \citep{Nesvorny_etal_2011JFC}. By not including meteoroid streams \citep[e.g.,][]{Christou_etal_2015} we neglect a fraction of meteoroids impacting Mercury's surface; however, these streams are temporarily and spatially concentrated with regards to Mercury's orbit, hence they likely do not control the global trends of its dust environment. At Earth, meteor showers are responsible for less than 10\% of the total flux of meteors observed by radars \citep{Brown_etal_2008}. 

Preliminary tests of our models showed that the characteristics (i.e. direction, velocity and flux distribution) of the meteoroid environment for particles with diameters between $D=30~\mu$m and $D=200~\mu$m do not show significant changes and thus we focus on meteoroids with diameter $D=100~\mu$m originating from JFCs and HTCs. The collisional lifetime for all meteoroids follows the best fit from \citet{Nesvorny_etal_2011JFC} for JFC meteoroids and \citet{Pokorny_etal_2014} for HTC meteoroids. Recent works suggest that the ratio between short period (i.e. JFCs) and long period (i.e. HTCs and/or OCCs) cometary dust is between 7:1 to 15:1 \citep{Nesvorny_etal_2011JFC,Carrillo-Sanchez_etal_2016}, and thus in this work we used 10:1 as an average ratio.

Counting individual impacts of test particles on planets suffers from low statistics due to small collision probabilities between meteoroids and a planet. To overcome this limitation of previous works \citep{Borin_etal_2016a}, we used a method developed by \citet{Kessler_1981} to calculate the intrinsic collisional probability. This technique is one of several available methods for evaluation of the intrinsic collisional probability between objects in the Solar System \citep{Opik_1951,Wetherill_1967,Greenberg_1982,Pokorny_Vokrouhlicky_2013}. By assuming that the semimajor axis, eccentricity and inclination of the projectile is constant in time while the other orbital elements are allowed to precess, it is possible to calculate the time for which both projectile and target share the same volume and thus evaluate the collisional probability between them. 

\section{Results}

\subsection{Meteoroid Flux and Velocity}

\begin{figure}[h]
\centering
\includegraphics[angle=-90,width=0.9\textwidth]{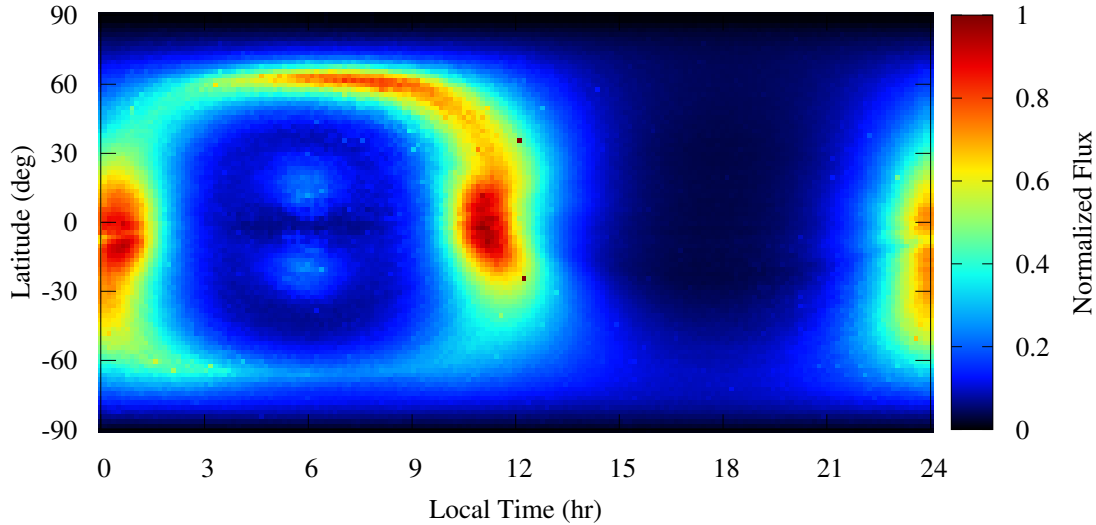}
%
%
\caption{Normalized radiant distribution of meteoroids impacting Mercury's surface averaged over its entire orbit. The x-axis represents the local time on Mercury and it is fixed with regards to sub-solar point (12h). Due to Mercury's eccentricity the location of the apex (approximately at 6h) changes along Mercury's orbit. The latitude is measured from Mercury's orbital plane (not the ecliptic 
).}
\label{Fig_1}
 \end{figure}

We first explore the radiant distribution of meteoroids impacting Mercury's surface averaged over a full orbital period. The radiant is the point in the sky from which the meteoroid appears to originate as seen from the planet's reference frame. Figure \ref{Fig_1} shows the normalized radiant distribution of meteoroids as a function of Local Time (LT) and latitude. Comparing the results displayed in Fig. \ref{Fig_1} with the observed radiant distribution on the Earth from the northern hemisphere \citep{Campbell-Brown_2008} and the southern hemisphere \citep{Janches_etal_2015,Pokorny_etal_2017} radar surveys, we readily identify all apparent sporadic meteoroid sources. Our model predicts that most of the meteoroid flux on Mercury comes from the ring structure composed from the helion source (HE, centered at $12\mathrm{h},0^\circ$), anti-helion source (AH, centered at $0\mathrm{h},0^\circ$), and the north/south toroidal sources (NT/ST, centered at around $6\mathrm{h},\pm 60^\circ$). The apex source (AP, centered at $6\mathrm{h},0^\circ$) is less prominent than at Earth. 
The difference between the results presented in Fig. \ref{Fig_1} and the distributions of meteor radiants observed by radars at Earth stems from the observational bias posed by the Earth's atmosphere and the radar detection limits. 
While smaller and slower meteoroids (i.e. meteors observed at the HE/AH sources) do not ablate efficiently in Earth's atmosphere and may not enough ionization to be observed by radars, meteoroids on Mercury impact the surface regardless of their mass or speed.
Figure \ref{Fig_1} clearly demonstrates the observed dawn-dusk asymmetry in Mercury's exosphere \citep{Burger_etal_2014} because there are virtually no impacts at around 18h. The same deficiency is observed at the Earth and is caused by the impact geometry, where the planet's orbital velocity prevents meteoroids from impacting from the anti-apex direction.

\begin{figure}[h]
\centering
\includegraphics[angle=-90,width=0.9\textwidth]{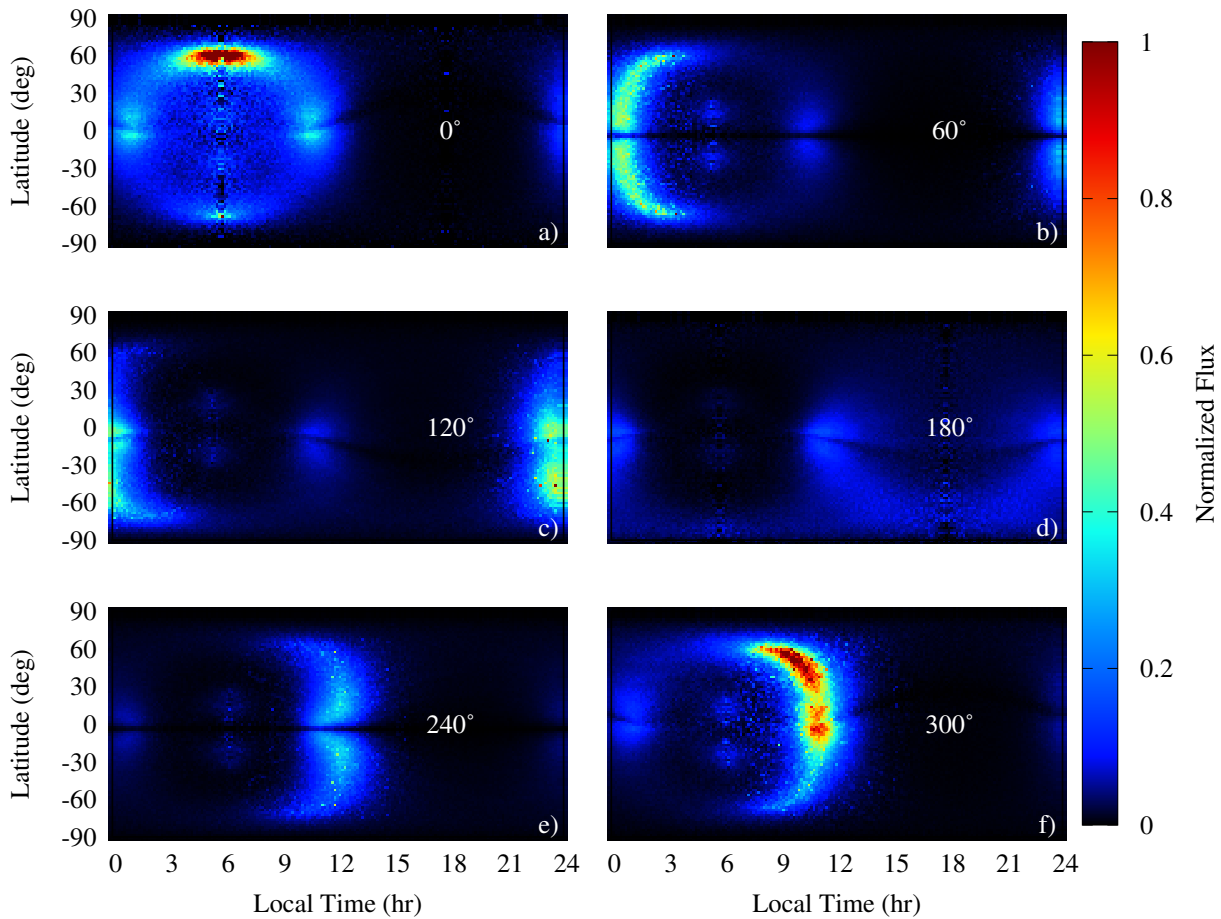}
%
%
\caption{The same as Fig. \ref{Fig_1}. but now for six different TAA (white number at $18\mathrm{h},0^\circ$ in all panels) along Mercury's orbit. The flux normalization is the same for all panels allowing to assess the relative flux changes along Mercury's orbit. 
}
\label{Fig_2}
 \end{figure}

Figure \ref{Fig_2} displays the seasonal variability of Mercury's meteoroid environment at six different phases of Mercury's orbit. 
At Mercury's perihelion (Fig. \ref{Fig_2}a; $\mathrm{TAA}=0^\circ$) the meteoroid environment is symmetric around the apex ($6\mathrm{h},0^\circ$) with the HE/AH sources exhibiting the same strength, the NT source dominating the total influx of meteoroids while the ST source is considerably weaker.
The NT/ST asymmetry stems from the non-zero inclination of Mercury's orbit with respect to the ecliptic. The results are remarkably different for $\mathrm{TAA}=60^\circ$, where the AH source becomes significantly stronger than the HE source (Fig. \ref{Fig_2}b). This is due to the non-zero eccentricity of Mercury's orbit causing the planet's velocity vector to be pointing away from the Sun, thus meteoroids originating from the AH direction will have a higher relative velocity as seen from Mercury than those coming from the HE direction. This effect becomes more striking for the case of $\mathrm{TAA}=120^\circ$ (Fig. \ref{Fig_2}c), where the additional movement of Mercury's velocity vector away from the Sun causes the shift of the AP and AH sources towards midnight. At aphelion ($\mathrm{TAA}=180^\circ$), Mercury's velocity vector is perpendicular to the direction of the Sun, and the planet's orbital velocity is minimal. This produces three effects on the dust precipitation: first the AH/HE sources have similar strengths, second the overall flux of meteoroids is much smaller, compared to the perihelion passage, third the meteoroids are able to hit Mercury from the anti-apex direction (Fig. \ref{Fig_2}d). Once the planet moves past its aphelion the HE source starts dominating the overall flux (Fig. \ref{Fig_2}e; $\mathrm{TAA}=240^\circ$). Closer to perihelion ($\mathrm{TAA}=300^\circ$) the HE and NT sources are dominating the total flux of meteoroids (Fig. \ref{Fig_2}f).  

\begin{figure}[h]
\centering
\includegraphics[angle=-90,width=0.9\textwidth]{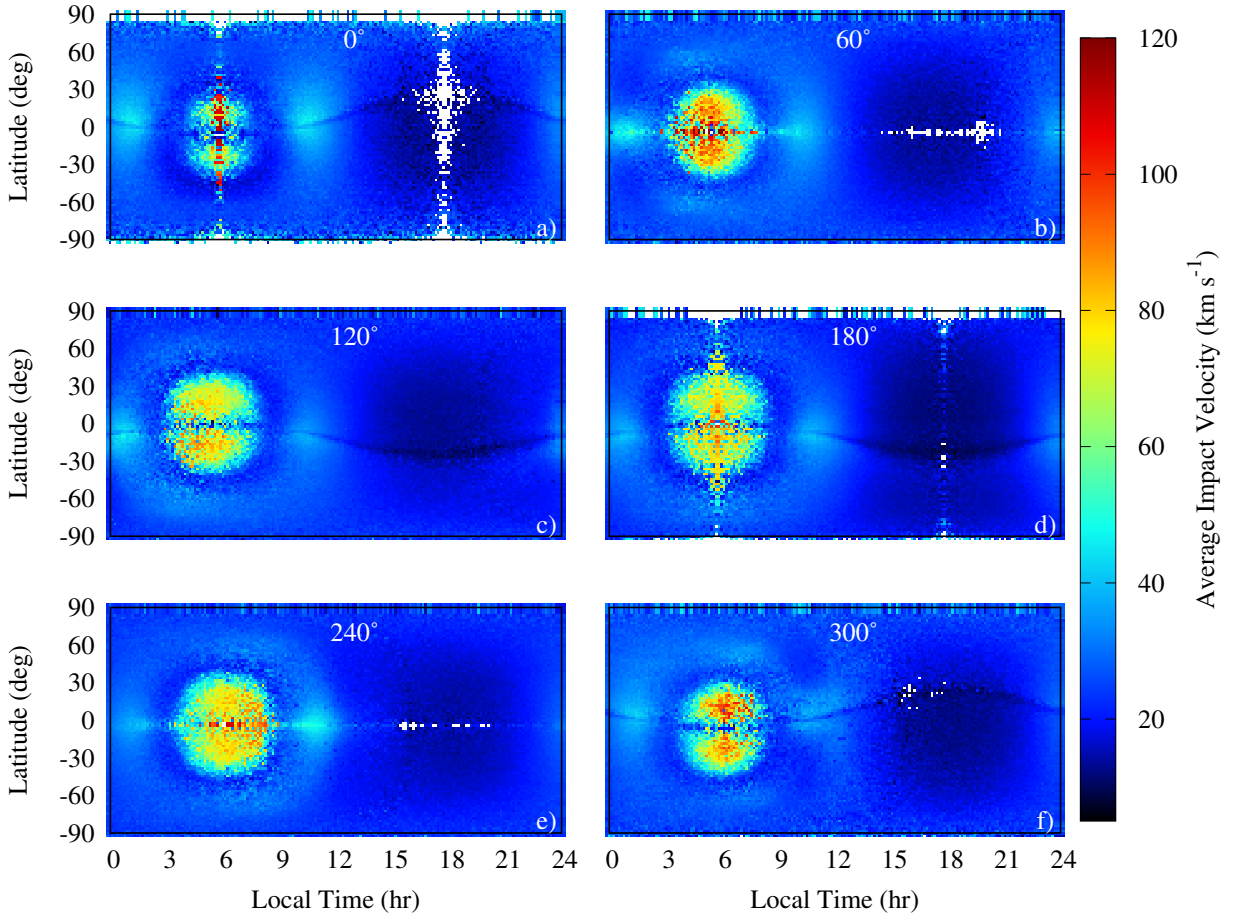}
%
%
\caption{The same as Fig. \ref{Fig_2} but now showing the average impact velocity $V_\mathrm{avg}$ in km~s$^{-1}$ for all radiant locations on the sky. TAA for each particular panel is represented by the white label at ($13\mathrm{h},80^\circ$). White colored radiants represent areas where no impacts were recorded.
}
\label{Fig_3}
 \end{figure}

Although plotting the flux from different radiant locations is valuable, to assess the effects of impacts on Mercury's surface, it is more insightful to study them relative to the meteoroids impact velocity. Figure \ref{Fig_3} shows the average impact velocity $V_\mathrm{avg}$ in km~s$^{-1}$ for six different times along Mercury's orbit. The most striking fact is that $V_\mathrm{avg}$ at the AP source is around 110 km~s$^{-1}$, while HE/AH meteoroids impact Mercury's surface at speeds 3 times lower ($V_\mathrm{avg}\sim$40~km~s$^{-1}$). This is because the AP source is populated by retrograde meteoroids, thus the AP source impact speed benefits from Mercury's high orbital velocity. While the flux distribution significantly changes along Mercury's orbit, the $V_\mathrm{avg}$ is modulated by Mercury's orbital velocity ($V_\mathrm{orb}(\mathrm{TAA}=0^\circ)=58.96$ km~s$^{-1}$, and $V_\mathrm{orb}(\mathrm{TAA}=180^\circ)=38.86$ km~s$^{-1}$). The AP source reaches the maximum $V_\mathrm{avg} \sim 110$ km~s$^{-1}$ at perihelion ($\mathrm{TAA}=0^\circ$), while the minimum $V_\mathrm{avg}\sim80~\mathrm{km~s^{-1}}$ occurs during aphelion passage ($\mathrm{TAA}=180^\circ$). HE/AH source velocities also peak at perihelion $V_\mathrm{avg}\sim{}40$ km~s$^{-1}$, the difference in $V_\mathrm{avg}$ between perihelion and aphelion passage is not as pronounced as for the AP source, $V_\mathrm{avg}\sim{}35 \mathrm{km~s}^{-1}$ at aphelion.

\subsection{Surface Response to Meteoroids - Impact Vapor Rate}
The relative flux of meteoroids and their impact velocities allow us to estimate the relative impact vapor rate $\mathcal{V}$ from each surface element of the planet. Average impact velocities per radiant location, shown in Fig. \ref{Fig_3}, show significantly higher values than those reported in \citet{Cintala_1992, Marchi_etal_2005} and~\citet{Borin_etal_2016a,Borin_etal_2016b}. This is because, until now, authors studying impacts on Mercury only focused on low velocity meteoroid populations released from asteroids and/or short period comets. Omitting long period comets (HTCs in our case) leads to an  under-representation of orbits with high inclinations and, most importantly, retrograde orbits impacting from the apex direction. While the flux from the AP source might seem negligible compared to other sources, extreme impact velocities intrinsic to the AP source play a crucial role in shaping the impact vaporization of Mercury's surface.

To estimate $\mathcal{V}$ from each meteoroid we followed the approach of \citet{Cintala_1992} 
and used a simple parametric estimate described by
\begin{equation}
\mathcal{V} = V_p (c+dv+ev^2),
\end{equation}
where $V_p$ is the volume of the incident meteoroid, $v$ is its impact velocity, and $c,d,e$ are constants. We adopted the following values: $c=-3.33$, $d=-0.0102$, and $e=0.0604$, corresponding to vaporization of iron at 400K \citep[Table 3. in ][]{Cintala_1992}. 
Knowing the direction and impact velocity of every meteoroid we calculate impact vaporization maps of the whole Mercury's surface. 
We determine the contribution of each meteoroid to all surface elements by calculating the zenith angle of each meteoroid $\theta_M$ with regards to each surface element using the following expression
\begin{equation}
\cos{\theta_M} = \cos{\Delta h} \cos{\Delta \Phi},
\end{equation}
where $\Delta h$ is the difference between the hour angle of the surface element and the hour angle of the meteoroid, and $\Delta \Phi$ is the difference between the latitude of the surface element and the latitude of the meteoroid. The expression for $\theta_M$ is simplified for Mercury since its axial tilt is almost zero.

\begin{figure}[h]
\centering
\includegraphics[angle=-90,width=0.9\textwidth]{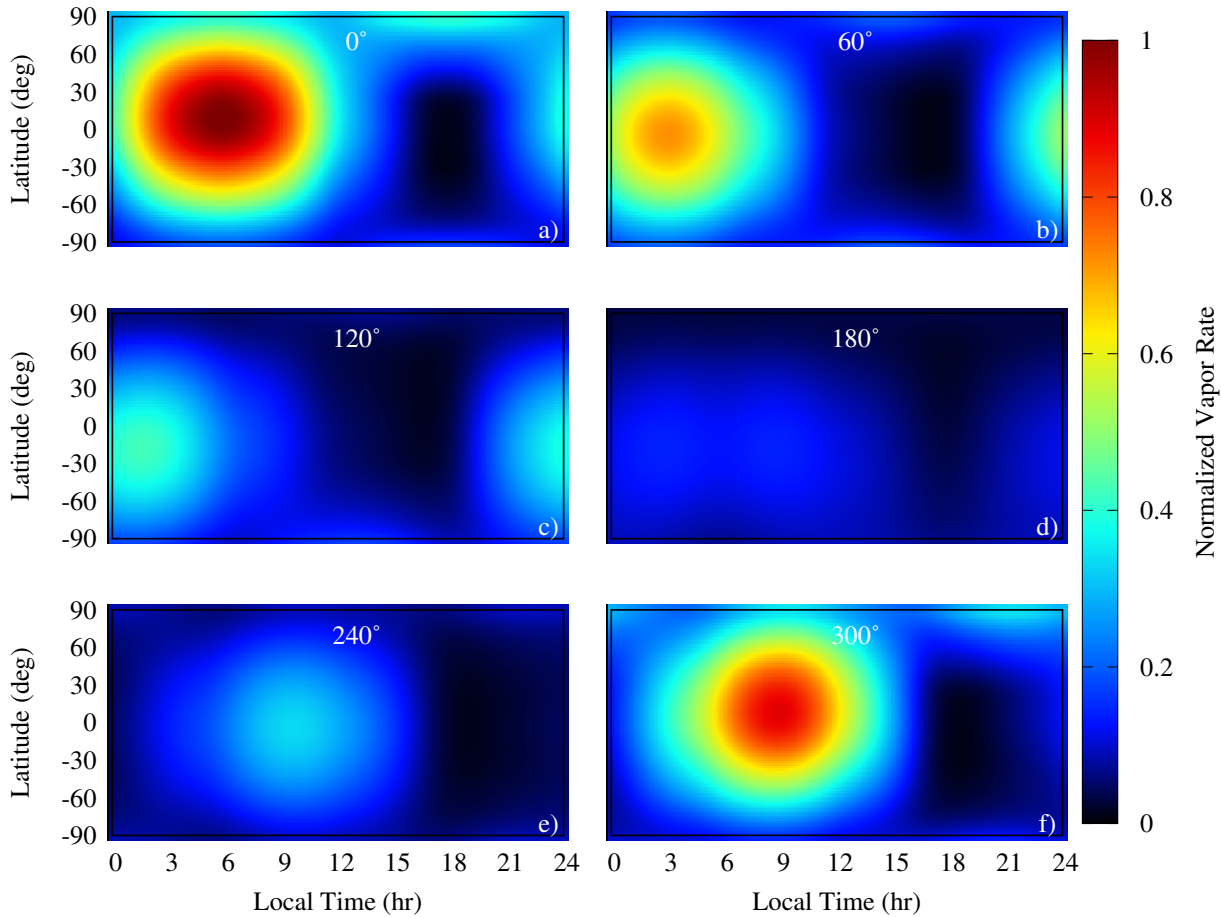}
\caption{The vaporization rate of Mercury's soil is non-uniform and changes when Mercury is moving away ($0-180^\circ~\mathrm{TAA}$) or towards (180-360) the Sun. TAA for each particular panel is represented by the white label at ($13\mathrm{h},80^\circ$).
}
\label{Fig_4}
 \end{figure}

\begin{figure}[h]
\centering
\includegraphics[angle=-90,width=0.9\textwidth]{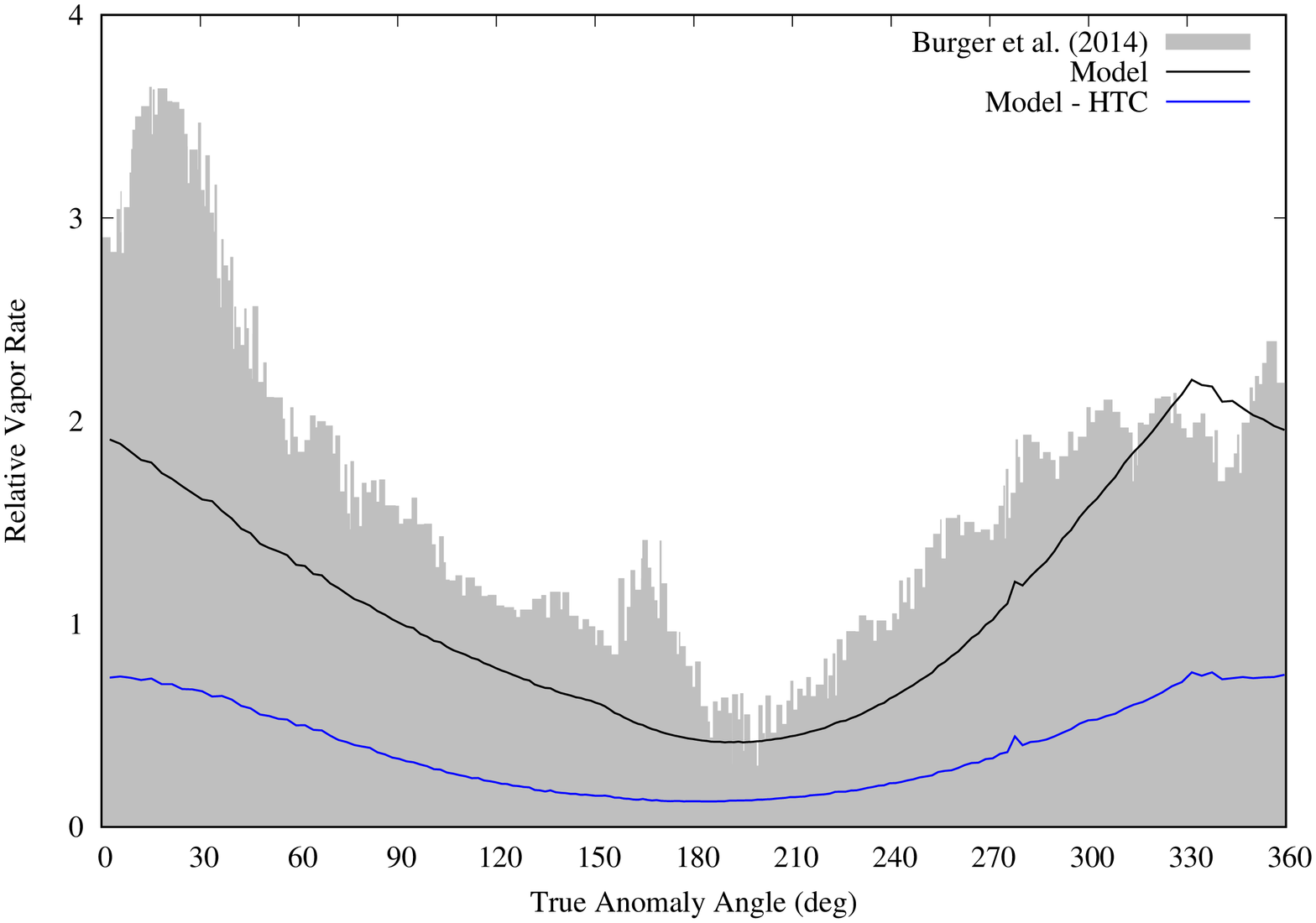}
\caption{Seasonal variations of the relative vaporization rate from our dynamical model (solid black line) compared to measurements of exospheric abundance of Ca from \citet{Burger_etal_2014} (grey columns). Contribution of HTC meteoroids is represented by the solid blue line.
}
\label{Fig_5}
 \end{figure}

Figure \ref{Fig_4} shows the normalized $\mathcal{V}$ for the whole surface at six different phases of Mercury's orbit. Comparison with Fig. \ref{Fig_2} emphasizes the importance the impact velocity for different radiant locations. At perihelion  ($\mathrm{TAA}=0^\circ$) 
the flux of meteors is mostly concentrated in HE/AH and NT sources, whereas the vaporization flux is centered at $(6h, 15^\circ)$. This is because meteoroids from the AP source impact Mercury surface at speeds 3 times higher that those from the HE/AH directions, leading to an order of magnitude higher $\mathcal{V}$. As Mercury moves from its perihelion towards its aphelion $\mathcal{V}$ follows the influx variations shown in Fig. \ref{Fig_2} and drifts in LT to the nightside  (Figs. \ref{Fig_4}b and \ref{Fig_4}c). At aphelion the difference between velocities of meteoroids from the AP source compared to those from the HE/AH sources is smaller than at perihelion and thus $\mathcal{V}$ is similar for all these sources (Fig. \ref{Fig_4}d). After the aphelion passage, LT during which maximum $\mathcal{V}$ occurs is shifted towards the dayside and the total vaporization rate increases significantly (Figs. \ref{Fig_4}e and \ref{Fig_4}f). At $\mathrm{TAA}=300^\circ$ the vaporization is shifted almost to 9h.

Finally, we show how well our modeled sporadic meteoroid environment at Mercury reproduces some of the observed features of Mercury's exosphere by MESSENGER measurements~\citep{Burger_etal_2014}. The number of emitted Ca atoms was derived by Monte Carlo models of particle transport fitted to the limb scans provided by the UVVS observations. \citet{Burger_etal_2014} parametrized the source of Ca atoms at the surface,  assumed a source centered at dawn (6h, $0^\circ$) that falls off exponentially with different characteristic widths, and showed that this source fitted the data for widths of $\sigma=30-40^\circ$. Our findings suggest that the source may not always be located near the dawn equatorial point (Fig. \ref{Fig_4}). Such motion of the source with Mercury's TAA can be tested with MESSENGER observations. In fact, Mg is inferred to peak during the morning hours, $8-10\mathrm{~AM}$ LT \citep{Merkel_etal_2017}. We replicate the Mg density peak displacement from the dawn terminator and attribute it to impacts. 

We compare results from our dynamical model with exospheric Ca rates derived by \citet{Burger_etal_2014} in Fig. \ref{Fig_5}. Seasonal variations of $\mathcal{V}$ from the sporadic meteoroid background agree reasonably well with MESSENGER measurements. This is an improvement of the dynamical model of \citet{Borin_etal_2016a}  where the influx rate is  essentially constant along Mercury's orbit. Enhancements observed around $\mathrm{TAA}=20^\circ$ and $\mathrm{TAA}=170^\circ$ cannot be reproduced by the sporadic background and require an additional significant source of meteoroids, where the most probable candidate seems to be 2P/Encke \citep{Killen_Hahn_2015,Christou_etal_2015}. 

\section{Discussion - Model Uncertainties}
This work presents initial results on the impact vaporization rates $\mathcal{V}$ produced by meteoroids with $D=100~\mu\mathrm{m}$. We calibrated the ratio of JFC and HTC populations at Earth with current estimates from \cite{Nesvorny_etal_2011JFC} and \citet{Carrillo-Sanchez_etal_2016}.
Dynamical meteoroid models have numerous free parameters characterizing the parent populations, the dynamical behavior in the Solar System, and the physical characteristics of meteoroids. Our experience suggests that the most important parameters shaping the meteoroid environment at Mercury are: a) the collisional lifetime of meteoroids, b) the ratio between JFCs and HTCs, c) the currently undetermined contribution of OCCs, and d) the size-frequency distribution (SFD) at the source populations. \citet{Pokorny_etal_2014} showed the effect that varying the collisional lifetime has on the arrival of HTC meteoroids at the Earth. From that work we assume that if the collisional lifetime is short \citep[comparable to ][]{Grun_etal_1985}, a vast majority of meteoroids will be collisionally destroyed before reaching Mercury's orbit. Secondly, the uncertainty in estimates of ratios between different populations of meteoroids impacting the Earth \citep{Carrillo-Sanchez_etal_2016} may play a major role for $\mathcal{V}$ of Mercury's surface. The high impact velocities from the apex source fed by retrograde meteoroids imply that significant changes in the distribution of $\mathcal{V}$ will be affected even by a small change in the flux density of HTC/OCC meteoroids at Mercury. And finally, the SFD influence on Mercury's meteoroid environment is mediated by the collisional lifetime. Collisions affect larger dust grains that undergo longer dynamical evolution before reaching Mercury's orbit. Thus, if the source SFD suggests that the majority of mass for one meteoroid population is weighted towards larger particles, such a population could ultimately cause less $\mathcal{V}$ on Mercury's surface than a population whose SFD is weighted towards smaller dust grains. A systematic assessment of these uncertainties on $\mathcal{V}$ patterns are beyond the scope of this manuscript and will be thoroughly analysed in our following work.

\section{Conclusion}
The combination of the dynamical models for JFCs and HTCs enables us to reproduce almost the entire observed meteoroid environment at Earth and its proximity, although OCCs may play a similar role to HTCs. By reproducing the dawn-dusk asymmetry of Mercury's exosphere and the seasonal variations (Fig. \ref{Fig_5}), we have shown that these models are also valid in the innermost regions of the Solar System. 

More importantly, we demonstrate for the first time the key role played by long period comets on the production of impact vapor. The previously omitted apex source consists of meteoroids with average velocities exceeding 110 km~s$^{-1}$ at perihelion and 85 km~s$^{-1}$ at aphelion. While the apex source flux is significantly lower than from other sources during the Mercury year, the high impact velocity of AP meteoroids results in a significant role in shaping the impact vaporization rate of Mercury's surface. 

The reduced complexity of atmospheric transport processes at Mercury compared to Earth, the collisionless nature of the exosphere, and the one-to-one correlation of refractory metals with the instantaneous source rate suggest that we can utilize MESSENGER measurements of Mercury's exosphere as a unique tool to constrain the dust population in the inner Solar System. For instance, if the motion of the atmospheric bulges in Mg and Ca away from the apex direction is less pronounced than shown here, this would imply an excess apex source because the movement of the source to the night/dayside is diminished by the stronger contribution of the apex source (Fig. \ref{Fig_4}). The significance of the apex source can be enhanced by considering meteoroids originating from Halley Type and Oort Cloud Comets. The trends which we predict here can also be applied to weathering processes and gardening rates over the last several million years because the velocity dependence of impact ejecta is similar to that for vapor. 

%

\acknowledgments
Authors would like to thank anonymous reviewers for their help. DJ/PP's works was supported with NASA's SSO and LDAP awards. MS acknowledges support from NASA grant NNX14AJ46G of the Planetary Atmospheres (PATM) Program.


\end{document}